\documentclass[11pt]{amsart}
\usepackage{geometry}                
\usepackage{graphicx}
\usepackage{amssymb}
\usepackage{epstopdf}
\usepackage{amsmath}
\usepackage{color}

\title{Prisoner's Dilemma on Real Social Networks: Revisited}
\author[Sharon M. Cameron and A. Cintr\'{o}n-Arias]{Sharon M. Cameron and A. Cintr\'{o}n-Arias}

\email{zsmc70@goldmail.etsu.edu}
\email{cintronarias@etsu.edu}
\address{East Tennessee State University \\
Department of Mathematics and Statistics\\
Box 70663\\
Johnson City, TN 37614-1701  
}

\date{August 27, 2012}            

\subjclass{Primary: 91A40; Secondary: 91D30}
\keywords{Game theory, social networks, small-world networks}

\thanks{S.M.C. received funding from NSF under grant DUE-0525447 and from ETSU Honors College under
a Research Discovery Position.  A.C.-A.
was funded by ETSU Presidential-Grant-in-Aid number E25150}
\begin{document}
\maketitle

\begin{abstract}
	Prisoner's Dilemma is a game theory model used to describe altruistic behavior seen in various populations. This 
	theoretical game is important in understanding why a seemingly selfish strategy does persist and spread throughout a 
	population that is mixing homogeneously at random. For a population with structure
	determined by social interactions, 
	Prisoner's Dilemma brings to light certain requirements for the 
	 altruistic strategy to become established. 
	Monte Carlo simulations of Prisoner's Dilemma are carried out using both simulated social networks and a dataset
	of a real social network.  In both scenarios we confirm the requirements for the persistence
	of altruism in the population.
\end{abstract}

\section{Introduction}

One well-documented example of altruism is 
food sharing.  Vampire bats ({\em Desmodus rotundus}) have nocturnal schedules,
they leave their roost for several hours during the night in search of prey they can feed from.
On following nights, it is customary for them to locate the prey they had previously fed on and continue 
their extraction of warm blood.  A handful of these bats may be unsuccessful in their food supply
search, however, they will not starve as 
some of their peers will regurgitate a portion
of the blood they acquired on that night, and share it with them.  Wilkinson \cite{wilk84} discovered that
sharing food by regurgitation, among wild vampire bats, is a function of reciprocation and it is
independent of the degree of relatedness.
In other words, as explained by Nowak \cite{nowaksup},
if a bat has previously fed another one, it is more likely this beneficiary re-pays the favor in the future.  Food sharing
is often cited as an example of direct reciprocity \cite{wilk84,nowaksup}, which is another way of referring to cooperation, the main
topic of this study.  Here we focus on mathematical modeling and simulation of cooperation.

The time evolution of cooperation is a subject of fascination for evolutionary biologists, that
finds it roots in the foundations of game theory \cite{neumort,maynard82,axelrod06,nowak06}.  Prisoner's Dilemma is perhaps one of the best-studied
theoretical games that describes altruistic behavior in organisms.  Typically, Prisoner's Dilemma
is formulated as a two-strategy and two-player game, where the payoffs are determined by years served in a jail sentence.
Indeed, the name of this game is coined from a scenario where two partners in crime are being held
for interrogation in separate rooms at police quarters, and they weigh in their options while the questioning takes place.

According to classical game theory, for populations of players that mixed homogeneously at random, 
cooperation is doomed to become extinct \cite{nowak06,easkle}.  
On other hand,
more recent developments suggest that when Prisoner's Dilemma is considered not just as a time-evolving process
but rather as a spatio-temporal evolving process, there are certain conditions that prevent the extinction
of cooperation \cite{ohtsuki}.  In this study we address space in the context of a social landscape for players 
of Prisoner's Dilemma.  We consider social networks with both simulated datasets 
and a dataset sampled from a collegiate social networking site.  \textcolor{black}{Specifically, 
our main contribution is to successfully validate a necessary condition for the establishment of 
cooperation (see \cite{ohtsuki} and references therein)
against an empirical dataset of a social network (friendship in a social networking site \cite{traud}). }

This paper is organized in the following way.  Versions of Prisoner's Dilemma
in well-mixed populations and those with network structure are introduced
in Sections \ref{wellmpd} and  \ref{pdnetwork}, respectively.  Models
for social networks with small-world properties are discussed in Section \ref{swnet}.
In Section  \ref{datafacebook} a dataset of a real social network is introduced.  A discussion
of the results is offered in the last section.

\section{Prisoner's Dilemma in Well-Mixed Populations}
\label{wellmpd}
In a well-mixed game, everyone is assumed to interact with one another, homogeneously at random.  Under this assumption,
cooperators may receive a benefit $b>0$ from other cooperating players, but cooperators also pay a cost $c>0$ for giving out benefits. 
Thus, the average payoff for cooperators is $b-c$.
On the other hand,
defectors, whom may only receive a benefit $b$ from cooperators, and whom neither pay a cost nor distribute any benefit,
end up having 
a payoff equal to $b$.  These payoffs are summarized into the strategy payoff matrix
	\(
	A= \begin{bmatrix}
		b-c & -c \\
		b & 0
	\end{bmatrix}.
\)

At time $t$ we have that $x_1(t) + x_2(t)=1$, where $x_1(t)$ denotes the density (fraction or proportion) of cooperators in the well-mixed 
population, while $x_2(t) = 1-x_1(t)$ denotes the density of defectors.  The fitness vector $f$ stores the expected fitness for 
each strategy (cooperation and defection), and
 results from the matrix-vector multiplication
\begin{equation}
	f = A x = \begin{bmatrix}
			b-c & -c \\
			b & 0
			\end{bmatrix}
			\begin{bmatrix}
				x_1 \\
				x_2
			\end{bmatrix}
			=
			\begin{bmatrix}
				(b-c)x_1 -c x_2 \\
				bx_1
			\end{bmatrix}.
\end{equation}
In other words, the fitness of the strategy cooperation is $f_1 = (b-c)x_1 -c x_2$, while
the fitness of the strategy defection $f_2 = bx_1$.  By defining the average fitness as follows
$\phi = x_1f_1 + x_2 f_2$ we can write the replicator equations for the Prisoner's Dilemma game \cite{nowak06}:
\begin{align}
	\label{creq}& \frac{dx_1}{dt} = x_1 \left(f_1 - \phi\right) =-cx_1x_2 \\
	\label{dreq}& \frac{dx_2}{dt} = x_2 \left(f_2 - \phi\right) = cx_1x_2
\end{align}
This system supports a stable equilibrium, such that 
$(x_1,x_2)\rightarrow(0,1)$.  
To see why it is stable we
reduce the system to one equation with a simple substitution, $u = 1-x_1$.  The reduced
replicator equation becomes $du/dt = -c u (1-u)=F(u)$.  Clearly, $u=0$ is a stable equilibrium
because $F'(0) = -c<0$ (by construction we assume a positive value for the cost, i.e., $c>0$).  This 
implies defection is the dominant strategy 
\textcolor{black}{(in fact, evolutionary stable strategy, see \cite{easkle,nowak06}) } , 
in the sense that cooperators go extinct, while defectors become established, taking over the entire well-mixed
population.  In other words, {\em natural selection favors defectors over cooperators} \cite{nowak06}.

\begin{figure}[h]
\begin{centering}
\includegraphics[height=2.3in,width=4in]{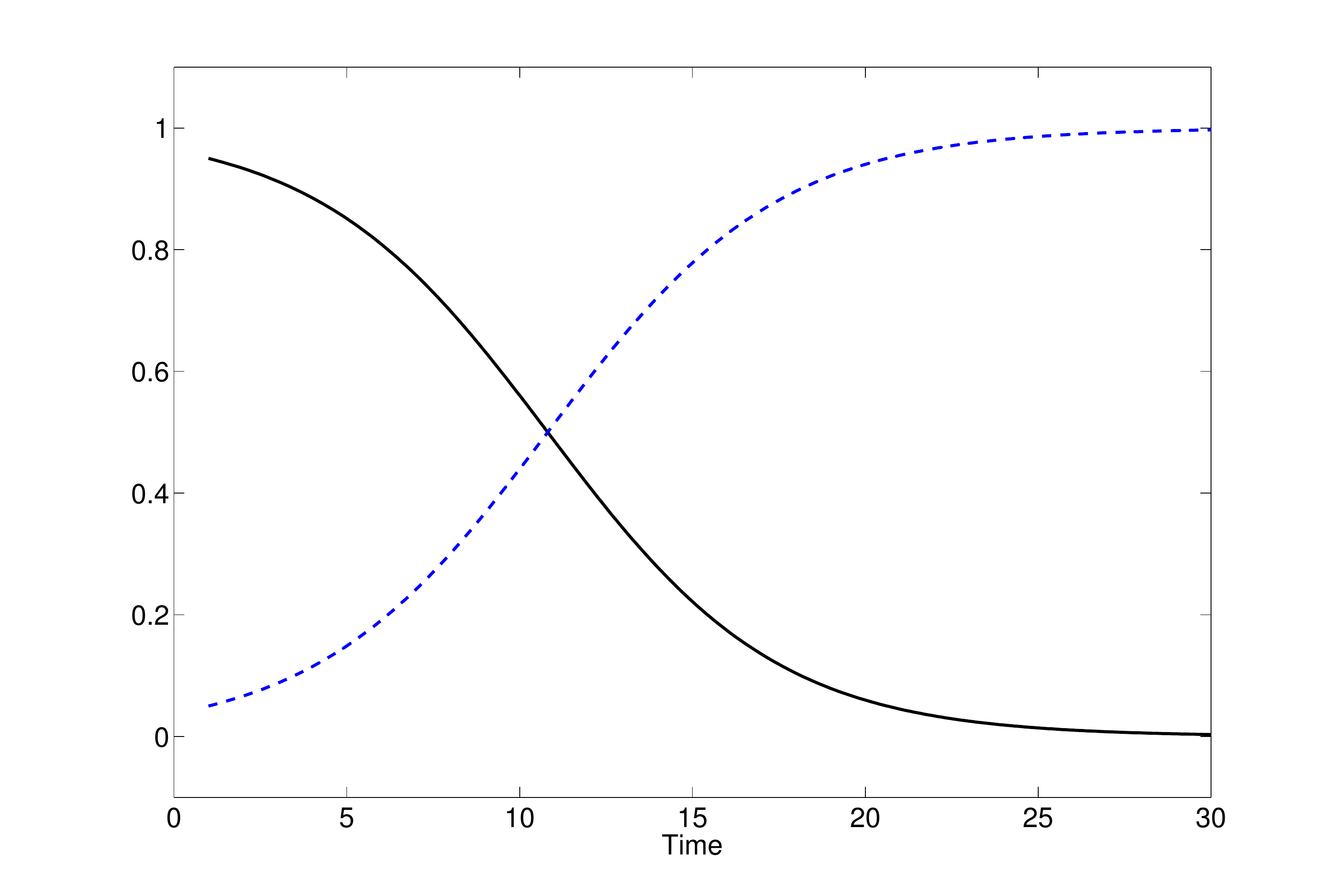}
\end{centering}
\caption{Numerical solutions to the replicator equations for the Prisoner's Dilemma.  Densities of cooperators $x_1(t)$ (solid curve) 
and defectors $x_2(t)$ (dashed curve) are displayed versus time $t$.
 Initial conditions and parameter values: 
 $x_1(0)=0.95$, $x_2(0)=0.05$, 
 $b=1.8$ and $c=0.3$.
}
\label{wellmixsol}
\end{figure}


Figure \ref{wellmixsol} illustrates the time evolution of cooperator and defector densities, that is,
numerical solutions to equations (\ref{creq}) and (\ref{dreq}).  The solid curve depicts the fraction 
of cooperators $x_1(t)$, while the dashed curve
 represents the density of defectors $x_2(t)$.
 For this particular numerical solution, we set $b = 1.8$ and $c = 0.3$. Also, 
 we started this simulation with 95 \% of cooperators and only 5 \% defectors. As can be seen, 
 the density of cooperators approaches zero as time progresses (i.e., $(x_1(t),x_2(t))\rightarrow(0,1)$ as
 $t\rightarrow 30$).  Even though the ratio of defectors to cooperators was initially one-to-nineteen, 
  meaning that for every 1 defector 
 there were 19 cooperators, which gave 
 cooperators an extremely biased favor initially, we still see the defectors taking over the population 
 and driving cooperators to extinction.

\section{Prisoner's Dilemma in Social Networks}\label{pdnetwork}

In this section we consider
a population of individuals who may engage in a decision-making scheme
equivalent to Prisoner's Dilemma.  In fact, social connections
by means of acquaintanceship, friendship, or levels of influence 
that can factor in decision-making are modeled with an undirected graph (network) \footnote{The words
graph and network, vertex and node, and edge and link will be used interchangeably.}, where
each vertex (node) represents an individual and an edge (link) denotes potential social 
ties \cite{gran}.

A social network provides a landscape where each node 
plays one of two strategies, cooperation or defection, and at 
each time step nodes decide whether to switch to a new strategy
or keep playing the same.  The key for these decisions is the payoff
per-node, which is now a space and time dependent quantity.  All nodes 
connected to a node say $i$, form its neighborhood, say $\Omega_i$.  To
compute the payoff of a node one needs to account for all pair interactions
(cooperator-cooperator, cooperator-defector, defector-cooperator and defector-defector)
happening in the node's neighborhood.  The strategy played by node $i$
is denoted with a binary vector $v_i$ defined as
\[
	v_i = \left\{
		\begin{array}{ll}
			\begin{bmatrix}
				1 \\ 0
			\end{bmatrix} 
			& \mbox{if node $i$ is cooperator} \\ \\
			\begin{bmatrix}
				0 \\ 1	
			\end{bmatrix}
			& \mbox{if node $i$ is defector}.
		\end{array}
	\right.
\]

The payoff of node $i$ at time $t$ is given by 
\begin{equation}
\label{payoffspatial}
	P(i,t) = \sum_{j \in \Omega_i} {v_i}^TA v_j,
\end{equation}

where $A= \begin{bmatrix}
		b-c & -c \\
		b & 0
	\end{bmatrix}$
denotes the strategy payoff matrix.  \textcolor{black}{The fitness of a node is the payoff re-scaled by an intensity
of selection parameter $w$, such that $0<w<1$.  When $w\rightarrow 0$ there is weak selection, 
while $w\rightarrow 1$ denotes strong selection \cite{ohtsuki}.  Thus,
we say the fitness of node $i$ at time $t$ is defined as follows:
\[	
F(i,t) = 1-w + w P(i,t),
\]
where the functional form of $F$ is known as linear fitness (see \cite{hauert12,ohtsuki}). 
}


The time evolution of Prisoner's Dilemma in a social network of players is subject to an updating rule.
In this study we considered the so-called ``death-birth" updating \cite{ohtsuki}:  at each time step a node
is chosen uniformly at random (unbiased) to die and its neighbors compete proportional to their fitness.  
Once this dying node is determined it becomes temporarily empty. This action may also be seen not necessarily
as an actual death of that member of the social network, but rather as if that node becomes a free-agent
and is open to be persuaded into playing other strategies.  The neighbors of this empty node compete for it, meaning
that the persuasion is proportional to their fitness.  Fitness is computed for each node in the neighborhood of  
the empty node (the empty node has to be excluded from the neighborhoods of each of node linked to it because it
has no strategy for time being), then the aggregate fitness for each strategy is calculated.  By aggregate fitness we
mean the total fitness of nodes playing cooperation and that of those playing defection.  The empty node decides
which strategy to play in the next time step in proportion to the aggregate fitness of cooperation and defection.  
(See appendix for additional details in pseudo code form.)

The ratio of benefit to cost serves as a threshold quantity that determines persistence of 
cooperation.  When this ratio is compared to the average degree of the network, average number of 
edges per node, denoted by $\langle k \rangle$, 
one obtains that
\begin{equation} \label{bck}
	\frac{b}{c} > \langle k \rangle
\end{equation}
is a necessary condition for selection to favor cooperation.  This threshold result
is derived from combining pair approximations and diffusion approximations \cite{ohtsuki},
where the fixation probability of a strategy is calculated.  This latter quantity represents
the probability that a single player of a strategy (either cooperation or defection) which
starts in a uniformly at random position in the network (unbiased), then gives rise to a lineage
of players of the same strategy, invading the whole population (see supplemental materials
of \cite{ohtsuki}).

In contrast to the well-mixed case, where for any values $b,c>0$
one obtains that the density of defectors always approaches one, 
$x_2(t) \rightarrow 1$ as $t$ gets large, for populations with structure, 
such as those with social network ties, it is seen that
cooperation is not doomed to be outcompeted.  Clusters of cooperators
can persist, provided some conditions are satisfied (with death-birth update
using aggregate fitness per strategy and when $b/c >  \langle k \rangle$).
In this study we intend to illustrate this feature using both synthetic data and
a dataset of a real social network.  The former are generated using Watts-Strogatz
algorithm for small-world networks, to be discussed in the next section.

\section{Models of Social Networks: Small-World Phenomenon and Watts-Strogatz Network Model}
\label{swnet}

Imagine we consider the following conditions for an experiment
on a social network. Randomly selected {\em  seed} individuals are 
asked to forward a letter with the ultimate goal of reaching a target recipient who
resides in Sharon, Massachusetts.  Even though seed individuals are given 
the name, address, and occupation of the target person, they are required to only
pass the letter along to someone in their circle of acquaintances that they 
know by their first-name.  S. Milgram \cite{mil} was the designer of this experiment
which resulted in measuring the average number of intermediaries in
these forwarding-letter chains: on average it took six individuals from seed
to target for the letter to arrive in Sharon, MA (see \cite{mil} and
chapter 20 of \cite{easkle}).

\begin{figure}
	\begin{center}
		\includegraphics[height=2in,width=6in]{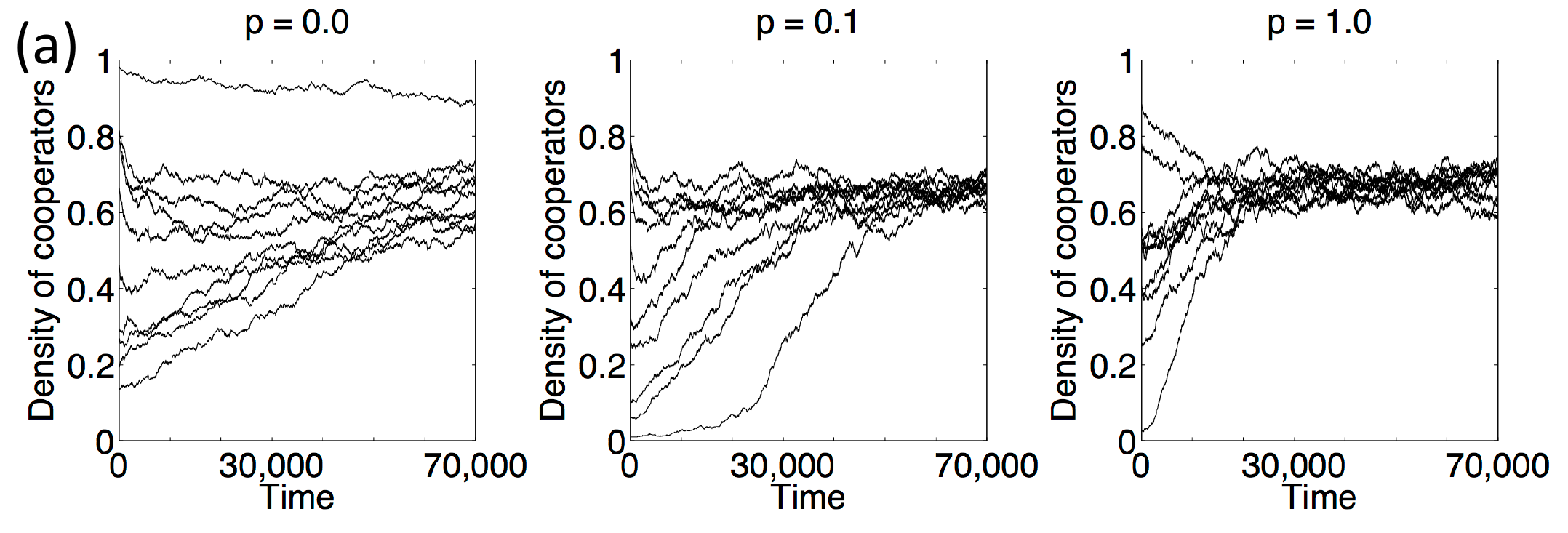}
		\includegraphics[height=2.5in,width=6in]{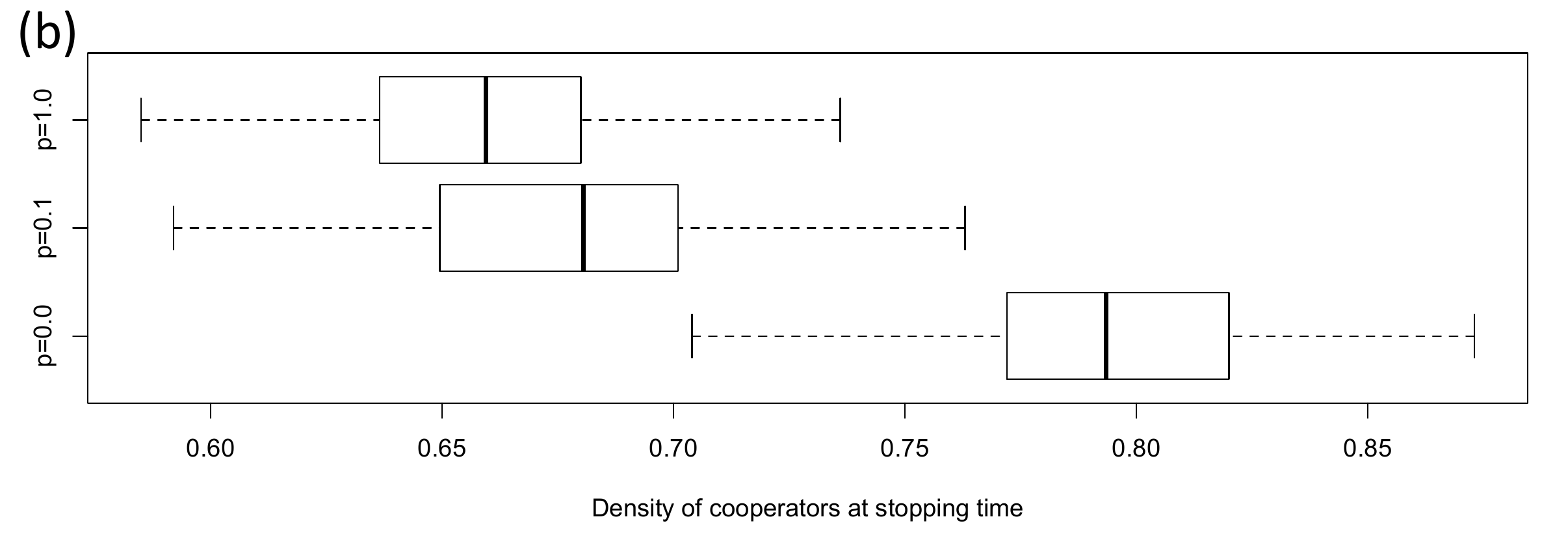}
		\includegraphics[height=2in,width=6in]{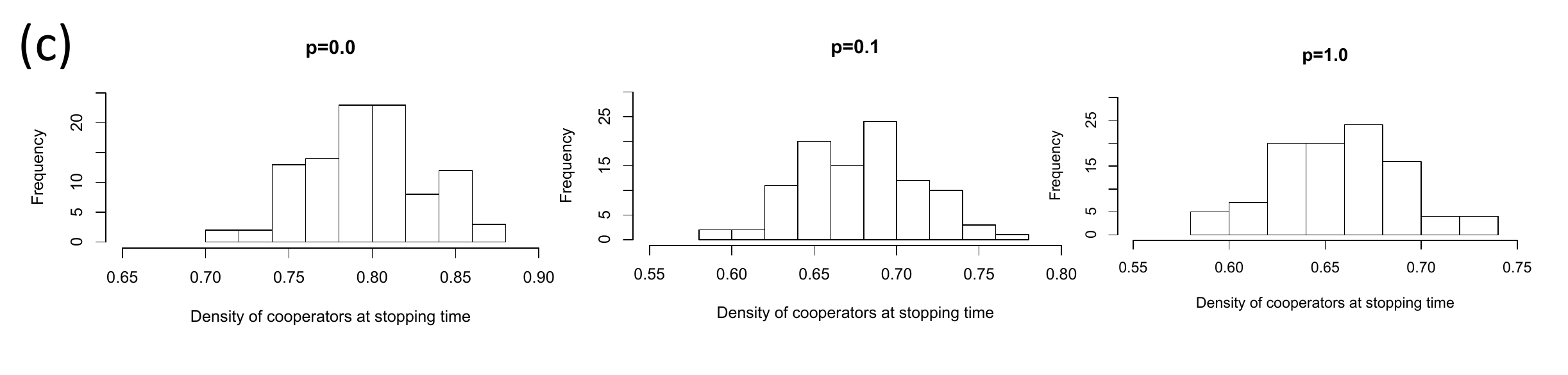}
	\end{center}
	\caption{Prisoner's Dilemma on small-world networks obtained with 
	rewiring probability values set to $p=0.0$, $p=0.1$, and $p=1.0$.
	Panel (a) displays the density of cooperators versus time, for 10 realizations with stopping time equal to $7.00\times10^{4}$.
	Panel (b) portrays boxplots, while Panel (c) displays frequency histograms, 
	of the density of cooperators with stopping time of $1.00\times 10^{6}$, for 100 realizations.
	Network and game parameter values: $n=1000$, $\langle k \rangle =4$, $b=1.8$, $c=0.3$ and $w=0.5$.
	}
	\label{wspd}
\end{figure}

This
became known as the ``small-world phenomenon" and it
speaks to structural properties of networks, where distance
between nodes is measured in terms of edges.  More precisely, 
paths are the concatenation of edges that connect a seed node
to a target node, the discovery of Milgram's experiment
would translate in saying that on average the forwarding-letter
paths consisted of six edges, indeed a short path \cite{mil,wattstro,watts99}.

Watts and Strogatz \cite{wattstro} proposed a model to construct 
families of networks with short paths, while also keeping track
 of an additional feature called clustering.  The latter refers
 to the existence of close triads or triangles, which
 denotes the ability of neighbors of neighbors to also be connected
 to each other by means of homophily (nodes connecting to other nodes
 who resemble themselves).  Watts-Strogatz network model transitions between two regimes:
 regular graphs, known to have high levels of clustering; and
 random graphs, known to have small characteristic path lengths.     
Gradual increments in the level of disorder are parametrized
by a tuning quantity: the probability of rewiring existing edges
in a network with a fixed number of nodes, denoted by $p$ where $0<p<1$.  Commonly, Watts-Strogatz networks
are referred to as small-world networks (additional details can be found in \cite{durr,easkle,newm,wattstro,watts99} and 
references therein).

 Our interest in the small-world networks relies in using them as a theoretical control group in the context
 of Prisoner's Dilemma.  More concretely, we are going to simulate Prisoner's Dilemma using networks
 generated with the Watts-Strogatz algorithm.  In this way, we simulate social influence by means of small-world
 networks while the theoretical game evolves in time.

\textcolor{black}{
{\bf Parameter values.}  Networks of size $n=1000$
and average degree $\langle k \rangle =4$ were employed.  The Prisoner's Dilemma parameter
values were chosen equal to those used in well-mixed populations for illustrations purposes (see Figure \ref{wellmixsol}):
$b=1.8$, $c=0.3$.  We decided to set the intensity of selection to a medium level ($w=0.5$) between 
strong ($w=1$) and weak ($w=0$) selection.  Rewiring probabilities were set at three different values: 
$p=0.0$ (regular graph), $p=0.1$ (graphs with large clustering coefficients and small characteristic
path length), and $p=1.0$ (random graphs).}

\textcolor{black}{
{\bf Initial conditions.}  Simple Random Sampling (SRS) was used to determine
initial conditions in the following sense.  Nodes were initially set to be cooperators
or defectors without preference (by means of SRS) due to degree, clustering, path length, or any other network attribute.
On the other hand, the number of nodes playing each strategy was chosen uniformly at random, with the only constraint
that total population remains constant at $n=1000$.
}

\textcolor{black}{
{\bf Stopping time and realizations.}  A stopping time of $T=1\times 10^{6}$ was employed.  A total of 100 stochastic realizations
of Prisoner's Dilemma were carried out for a fixed value of rewiring probability $p$. A network was drawn from Watts-Strogatz algorithm, 
with each fixed value of $p$, which was kept static during time steps $t=1$ through $t=T$ (i.e., over the course of one stochastic realization
of the theoretical game).
}

\textcolor{black}{
{\bf Update rule.}  A death-birth updating rule was implemented (see Section \ref{pdnetwork}), such that
$b/c$ exceeds the average degree:  $b/c= 1.8/0.3= 6$ and $ \langle k \rangle =4$.  This necessary condition
for the establishment of cooperation is precisely what we intend to validate with the present study.
}

\textcolor{black}{
Figure \ref{wspd}(a) displays a snapshot of the density of cooperators versus time, i.e.,
$x_1(t)$ versus $t$.  For the sake of resolution only 10 realizations are displayed with
time between $t=1$ and $t=70000$, where each discrete time step $t$ represents 
a round of the game being played.  Left-side, middle, and right-side figures in Figure \ref{wspd}(a) 
depict time series corresponding to $p=0.0$, $p=0.1$, and $p=1.0$, respectively.
}
\begin{table}[h]
	\caption{Mean and five-number summary of density of cooperators at stopping time, for 100 realizations
	of Prisoner's Dilemma on small-world networks (see caption of Figure \ref{wspd}).}
	\begin{center}
		\begin{tabular}{ccccccc} \hline \hline
			Rewiring & Minimum & First Quartile & Median & Mean & Third Quartile & Maximum \\ \hline\hline
			 $p=0.0000$  &0.7040  &0.7725  &0.7935   &0.7969  &0.8200  &0.8730   \\ \hline 
			 $p=0.1000$ & 0.5920 & 0.6498  &0.6805  &0.6785  &0.7010  &0.7630   \\ \hline 
			 $p=1.0000$& 0.5850   & 0.6368  &0.6595  &0.6580  &0.6800  &0.7360  \\ \hline 
		\end{tabular}
	\end{center}
	\label{ws5num}
\end{table}

\textcolor{black}{
Figures \ref{wspd}(b)--(c) summarize results of 100 realizations where the stopping time is 
$T=1\times 10^{6}$. In Figure \ref{wspd}(b) we find boxplots of $x_1(T)$,
while Figure \ref{wspd}(c) displays histograms of $x_1(T)$  for the three
values of rewiring under consideration.  Across these three types of rewiring
we observe a consistent unimodal shape of the distributions of 100 samples, where
no particular skewness is observed.  In fact, consistent with the lack of skewness
the variability in the samples does not seem to fluctuate drastically and no outliers are
included either.
}

\textcolor{black}{
	The mean along with the five-number summary of $x_1(T)$
are specified in Table \ref{ws5num}.  For each value of $p$, the mean
and median are fairly close to one another, such that they match when
round off to one decimal digit.  In a community with regular
graph structure, $p=0.0$, the median fraction of cooperators
at stopping time is 0.7935 with interquartile range (IQR) of 0.0475.
Simulated communities with high levels of clustering and
small average distance between nodes, $p=0.1$, 
exhibit a median of $x_1(T)$ equal to 0.6805 with IQR
equal to 0.0513.  On the other hand, in communities with 
simulated random graph structure, the median and IQR of 
$x_1(T)$ are  equal to 0.6595 and 0.0433, respectively.
The values of IQR in these three cases confirm what is observed
in Figure \ref{wspd}(b)--(c), i.e., no strong fluctuations in variability of $x_1(T)$
samples is noticed.
}

\textcolor{black}{
We now use the median to comment on average behavior of Prisoner's Dilemma time evolution,
among simulated social networks (communities).  It is well-known that Watts-Strogatz algorithm
provides families of networks at $p=0.1$ that have favorable local and global features.  At this value
of rewiring, networks have small average characteristic path lengths (global property) and
large clustering coefficients (local property) \cite{wattstro}.  The median of $x_1(T)$ drops substantially from $p=0.0$
to $p=0.1$ (see Figure \ref{wspd}(b) and Table \ref{ws5num}): a drop of basically $1.0\times 10^{-1}$.  On the other hand,
while the median of $x_1(T)$ decreases again from $p=0.1$ to $p=1.0$, it is not as drastically as in the previous case.
}

\textcolor{black}{ 
The comparison of Prisoner's Dilemma across two extremes of small-world networks,
top versus bottom bloxplot in Figure \ref{wspd}(b), suggests the coverage of cooperators
in the simulated communities drops from 80\% to 66\%.  In other words, structure
plays a role in the final number of cooperators at stopping time.  In a more general sense,
these boxplots in Figure \ref{wspd}(b) confirm that clusters of cooperators
persist in these simulated social networks over time.  The choice of stopping time 
at $T=1 \times 10^{6}$ guarantees a burn-in phase.  Longitudinal trends of $x_1(t)$ 
with $t$ exceeding $5\times 10^{5}$ (not displayed here) assure
a steady-state-like behavior.
}

A closer examination of the solid curve displayed in Figure \ref{wellmixsol}, along with
 the realizations of Figure \ref{wspd}(a),
leads to compare Prisoner's Dilemma on well-mixed communities versus small-world networks.  As it was
discussed in Section \ref{pdnetwork}, for social networks with $b/c > \langle k \rangle$
cooperators are not condemned to extinction, unlike in well-mixed populations.  The same
parameter values were used in the numerical solutions of Figure \ref{wellmixsol} and the simulations
of Figure \ref{wspd}(a):  $b=1.8$ and $c=0.3$.  We see in Figure \ref{wellmixsol}  
that after 25 rounds of the game, cooperators basically disappear in a well-mixed community, while
Figure \ref{wspd} illustrates a sustained persistence of cooperators over time.

\section{Dataset of a Social Network and Simulated Prisoner's Dilemma}
\label{datafacebook}

The first decade of the twenty-first century has seen the rise and establishment
of readily accessible technology to communicate with others simply by 
hitting a key stroke
in a mobile device, whether it is a laptop, a smartphone, or a tablet.  The World Wide Web continues
to host the so-called ``social networking sites" (SNS).  These are the up-to-date versions of 
forums that facilitate exchanges 
which are remarkably casual and informal, occurring remotely in real-time.  

According to Boyd and Ellison \cite{boyd08} SNS are web-based tools that 
accomplish three main objectives: (1) easy development of a profile
with the option of making it public; (2) intuitive interface for constructing
lists of users to connect with;
(3) access to lists of users sharing a connection.

Today, one of the well established SNS is Facebook\footnote{\texttt{http://www.facebook.com/}}, where
users easily share personal information by means of photos, videos, 
and email.  Facebook also facilitates surveying opinions on topics of specific interest
and it is known to even promote organization of events.  In the early days Facebook membership was
restricted to university affiliation.  In other words, it served as collegiate social networking site requiring
users to have a valid email with an edu-suffix.  It first launched at Harvard University in early 2004 and it gradually
expanded to other universities.  The email requirement made Facebook users feel exclusive
because they had membership to a private community \cite{boyd08}.  By September 2005,
Facebook moved forward to integrate professionals working within corporate networks and high school students.
However, Facebook did not allow its users to make their profiles public to all users right away.  This was
a substantial difference relative to other SNS \cite{boyd08}, and it meant that it preserved 
a strong sense of local community.
\begin{figure}[h]
	\begin{center}
		\includegraphics[width=1.7in,height=1.35in]{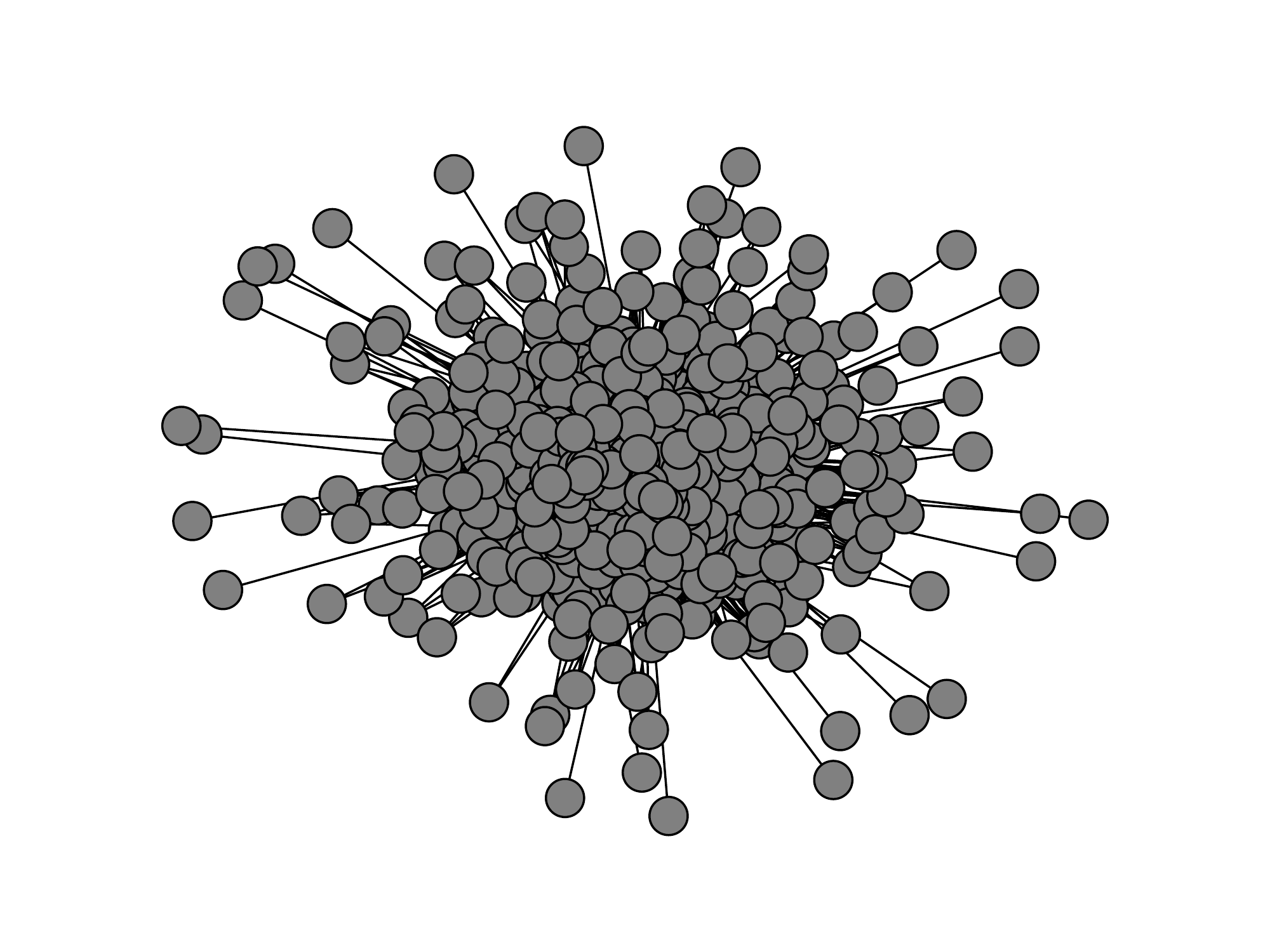}
	\end{center}
	\caption{Network visualization of the largest connected component
	 in a social network dataset sampled at the California Institute of Technology \cite{traud}.}
	\label{caltechfriends}
\end{figure}

A Facebook friendship between two users means there is a link connecting 
their profiles.  Moreover, for these links to be established Facebook requires confirmation
of a ``friendship request".  In this sense, Facebook friendships determine
a network of users, in so many words: a graph of undirected edges, where each node
represents a Facebook user.  For examples of social network analyses using this type
of datasets see \cite{lewis08,mayer08,traud12}.

The dataset employed here is a subset of those used by Traud, et al. \cite{traud}, it consists 
of a complete set of users and all the links between them occurring on September 2005 at
the California Institute of Technology.  Figure \ref{caltechfriends} displays a network visualization
of the Caltech dataset, where nodes and links denote Facebook members and friendships,
respectively.

In their comprehensive analysis, Traud, et al. \cite{traud} quantify some of the basic network
characteristics of the Caltech dataset.  For example, the network size is $n=1099$ with only
762 nodes belonging to the largest connected component.  Moreover, there are 16651 edges
 within the largest connected component.  The average degree is $\langle k \rangle =43.70$, while
 the mean clustering coefficient is 0.41.  Traud, et al. \cite{traud} point out that when
 comparing clustering, by two different measures, against the datasets of another four
 universities, the Caltech dataset has the largest clustering.  In their study, Traud, et al. \cite{traud},
one of their main goals is to detect significant clusters of nodes (community structure), by using
unbiased algorithms.  They find the Caltech dataset has 12 communities.  Using the Caltech dataset, 
we carried out Monte Carlo simulations of Prisoner's Dilemma and below we give details
of the implementation.

\begin{figure}[h]
\begin{centering}
\includegraphics[height=2in,width=4in]{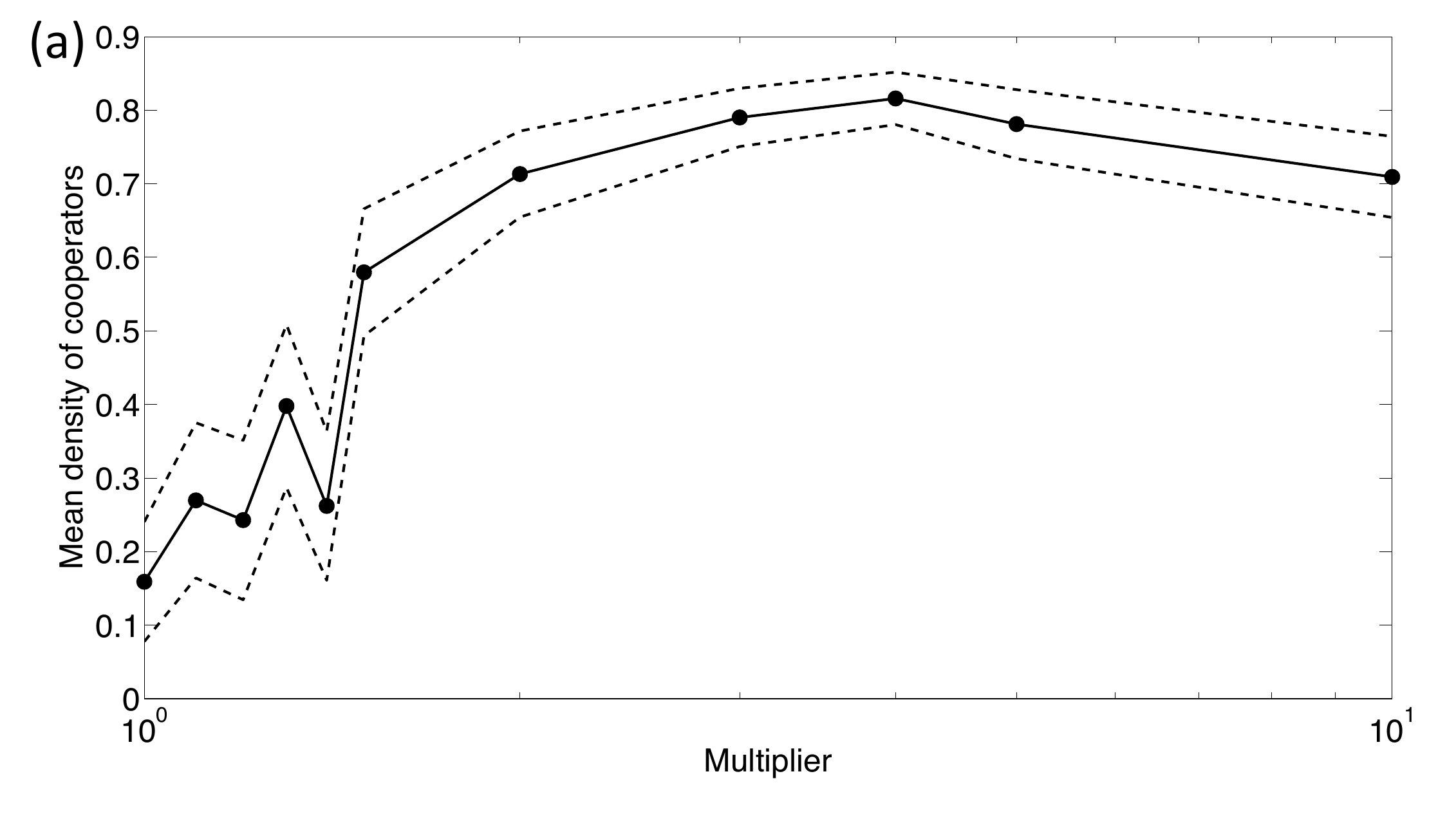}
\includegraphics[height=1.5in,width=5.5in]{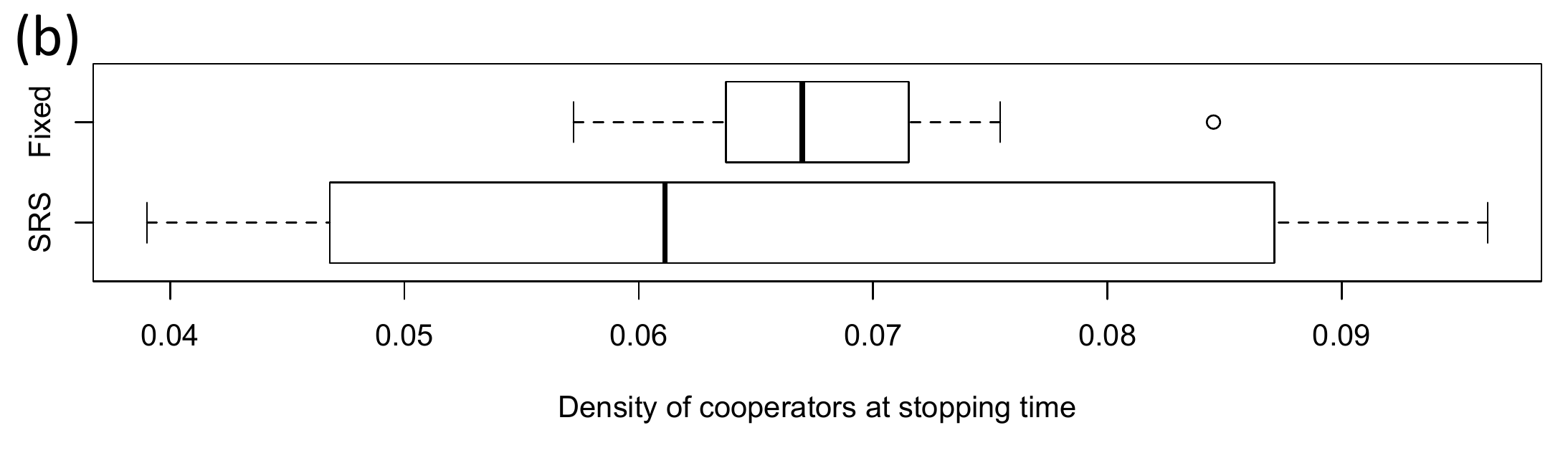}
\end{centering}
\caption{Diagnostic tests of Prisoner's Dilemma simulations on a dataset of a real social 
network \cite{traud}.  Panel (a) displays mean density of cooperators at stopping time versus
a multiplier $\beta$, where it is assumed $b/c = \beta \langle k \rangle$ for 
$\beta \in \{1.0,\ 1.1,\ 1.2,\ 1.3,\ 1.4,\ 1.5,\ 2.0,\ 3.0,\ 4.0,\ 5.0,\ 10.0\}$.  Dashed curve denotes variability:
the mean plus minus one standard error.  These averages were computed out of 10 realizations
with stopping time $T=1\times 10^{5}$. In Panel (b) boxplots of the density of cooperators at stopping time
are depicted.  The samples in these boxplots were obtained from 10 realizations with stopping time $T=1\times 10^{6}$
and $b/c = \langle k \rangle$.  Two types of initial conditions were tested:
Fixed initial conditions (top boxplot), where half of the population 
were initialized as cooperator while the other half were set as defectors; Initial conditions by simple random sampling (bottom boxplot),
where the initial number of cooperators was chosen uniformly at random between 1 and $n$.}
\label{diagnostics}
\end{figure}

{\bf Parameter values.} Because the necessary condition $b/c > \langle k \rangle $ is 
at the central stage of this study, we decided to explore the ratio $b/c$ as a linear function
of the average degree $ \langle k \rangle$.  In other words, for diagnostic tests we supposed
that $b/c = \beta \langle k \rangle$ for some $\beta\geq 1$, where $\beta = 1$ denotes a 
borderline case scenario.  Values of $\beta$ were considered in $[1.0,1.5]$ along with
values in $[2,10]$.  Setting initial conditions to 50\% cooperators and defectors at time 
$t=1$, and stopping time $T=1\times 10^{5}$, led to results displayed
in Figure \ref{diagnostics}(a) for 10 realizations.  It is seen in Figure \ref{diagnostics}(a)
that the mean of $x_1(T)$ is an increasing function of $\beta$, where $\beta\in[1,10]$.  
For $\beta =1$ it is seen that $x_1(T)$ is above 0.1 (at least 10\% of the network remains playing cooperation),
while for $\beta \geq 2$ then $x_1(T)$ is no less than 0.7 (more than 70\% of cooperators remain in the network).
Based on this diagnostic we opted to set $b/c = 3.41\langle k \rangle $ (a value of $\beta$ between 3 and 4):
more specifically, we set $b/c = (3.41)(44.0)\approx 150$, i.e., $b=150$, $c=1$.  The value of 
intensity of selection was set at $w=0.5$, halfway through weak and strong selection.

{\bf Initial conditions.}  The effect of two types of initial conditions was also vetted.  We consider
the borderline case $b/c = \langle k \rangle$ and set the stopping time as $T=1\times 10^{6}$.  Figure \ref{diagnostics}(b)
depicts results of 10 realizations.  The top boxplot of the samples of $x_1(T)$, corresponds to fixed initial conditions, i.e.,
where 50\% of the nodes were initially set to be cooperators.  On the other hand, the bottom
boxplot corresponds to initial conditions determined by Simple Random Sampling (SRS), where the initial number of cooperators
was chosen uniformly at random between 1 and $n=1099$.  The choice of which nodes were initially set to play cooperation
was made independently of any network attributes.  Comparison of the median in these boxplots displayed in Figure \ref{diagnostics}(b)  
suggests cooperators reach very low levels at $t=T$ (but yet they are not extinct, at least on average), something that is 
expected in the borderline case $b/c = \langle k \rangle$.  Even though fixed initial conditions exhibit an outlier for 
$x_1(T)$ samples, and some skewness, the variability remains substantially narrower in fixed versus SRS initial conditions.  
We opted for setting initial conditions by SRS to allow more variability in the simulations outcome.

{\bf Stopping time, realizations and updating rule.}  A death-birth updating rule was employed (Section \ref{pdnetwork}),
while the stopping time was set as $T=1\times 10^{6}$ and
100 realizations of Prisoner's Dilemma were carried out using the Caltech dataset. 

\begin{table}[h]
	\caption{Mean and five-number summary of density of cooperators at stopping time, obtained from Caltech dataset.}
	\begin{center}
		\begin{tabular}{cccccc} \hline \hline
		 	Minimum & First Quartile & Median & Mean & Third Quartile & Maximum \\ \hline\hline
			0.2185  &0.7055&  0.8381&  0.7881&  0.9038&  1.0000 \\ \hline
		\end{tabular}
	\end{center}
	\label{caltech5n}
\end{table}

\begin{figure}[h]
\begin{centering}
\includegraphics[height=1.82in,width=5in]{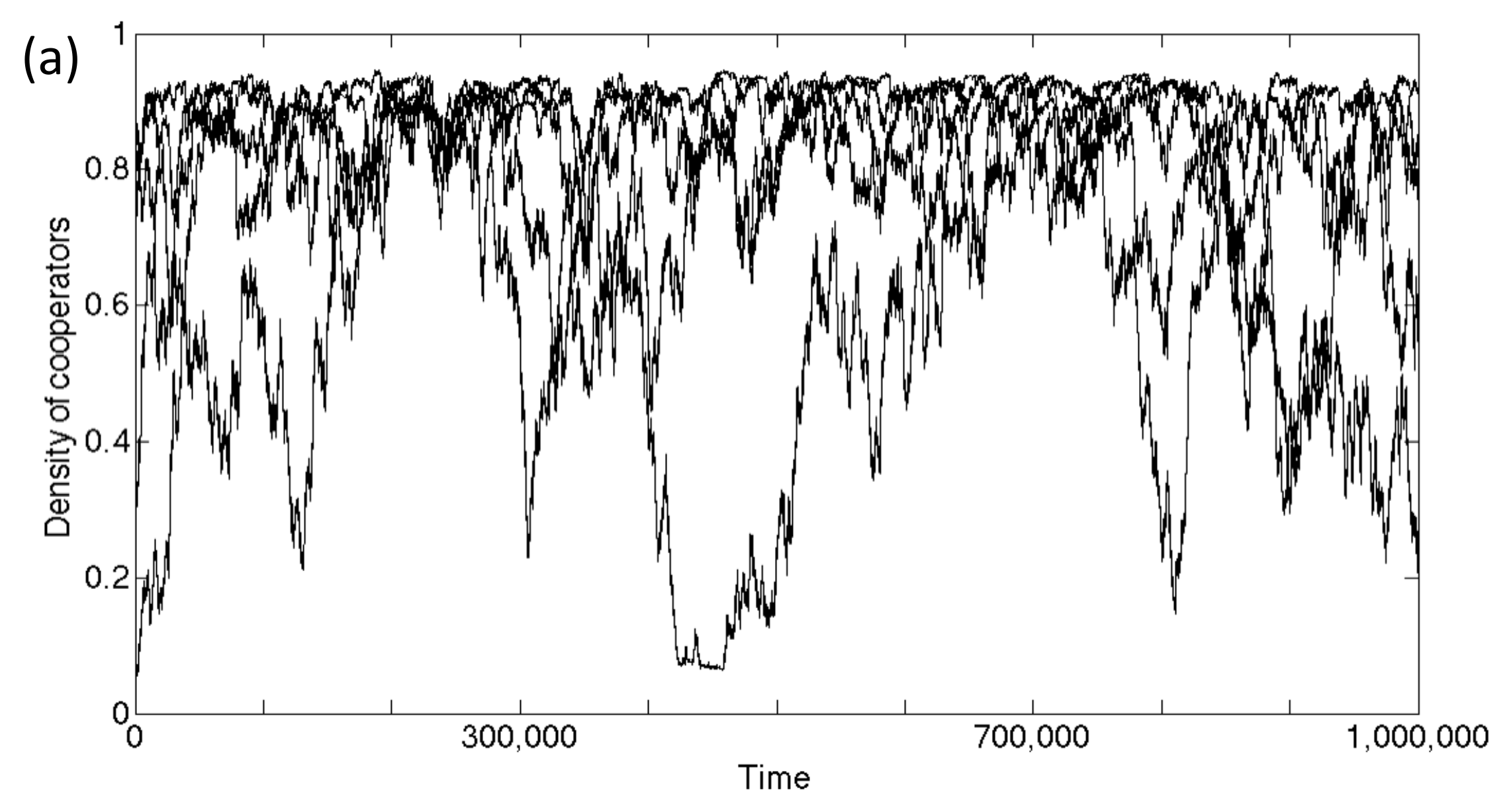}
\includegraphics[height=1.82in,width=5in]{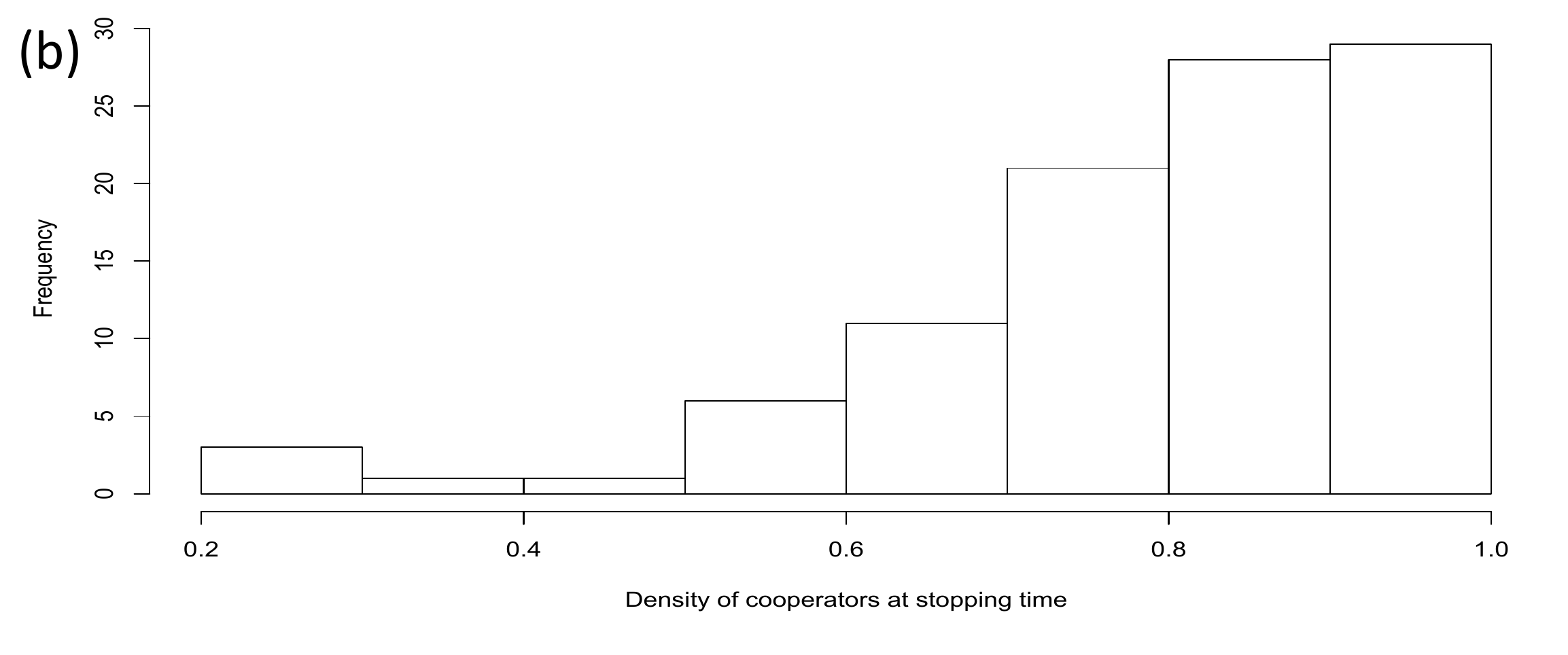}
\includegraphics[height=1.8in,width=5in]{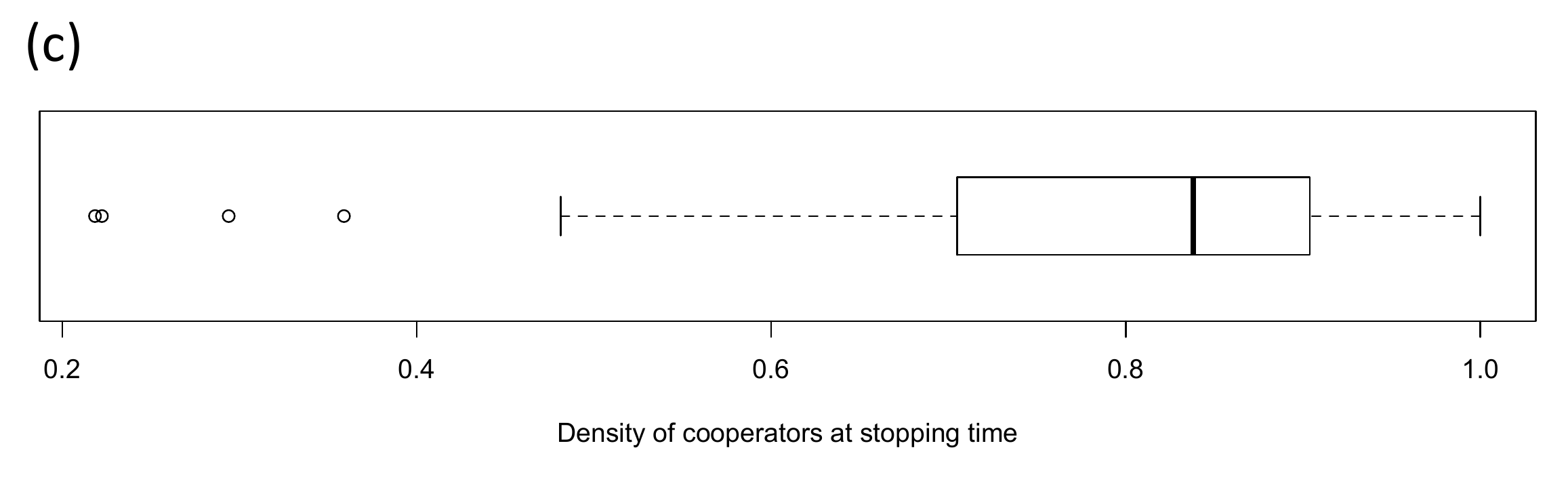}
\end{centering}
\caption{
Prisoner's Dilemma on a real social network, dataset sampled at California Institute of 
Technology \cite{traud}.  Panel (a) displays 10 realizations of the 
density of cooperators versus time with stopping time $T= 1\times10^{6}$.  Panel (b)
depicts the histrogram while Panel (c) displays the boxplot of samples of
cooperators density at stopping time, for 100 realizations of Prisoner's Dilemma.
SRS initial conditions were used.
Game parameter values: $b=150$, $c=1$ and $w=0.5$.  This dataset has size $n=1099$ and average degree 
$\langle k \rangle = 44.0$.  Game parameter values were chosen to ensure $b/c > \langle k \rangle $ \cite{ohtsuki}.
}
\label{coopcaltech}
\end{figure}

Panel (a) of Figure \ref{coopcaltech}
displays only 10 (out of the 100 realization) curves of cooperators density, for the sake of enhanced
resolution.  There is clear evidence of patterns supporting persistence of cooperation, 
as it is revealed in this subset of the 100 realizations.

Another observed feature in Panel (a) is downward-spike
 temporal patterns, for a handful of realizations.  In other words,
the density of cooperators in these cases drops remarkably, but it seems to return
back to sustained levels.  Similar patterns of drops in cooperation density
have been reported before by Egu\'{i}luz, et al. \cite{egui} (see Figure 5),
and by Kim, et al. \cite{kim02} (see Figure 3(b)), albeit with 
different versions of Prisoner's Dilemma.

The histogram of samples of cooperators density at stopping time $T=1\times 10^{6}$
is displayed in Figure \ref{coopcaltech}(b).  Considerable skewness is observed,
in comparison to small-world networks (see Figure \ref{wspd}(c)).  Moreover,
skewness is also confirmed by the boxplot in Figure \ref{coopcaltech}(c), 
where a handful of outliers appear.  The latter suggest low levels of sustained cooperation, but
no necessarily extinction. 

The five-number summary and mean of $x_1(T)$ are given in Table \ref{caltech5n}.
As expected because of skewness the mean (0.7881) and median (0.8381) are distant from
one another, relative to the simulations on small-world networks (see Table \ref{ws5num}).  Also,  
the IQR of the samples is 0.1983, implying IQR of the simulations with the Caltech dataset
is one order of magnitude larger than the IQR's obtained with small-world networks.

Since the median of $x_1(T)$ is 0.8381 one concludes that, 
on average, clusters of cooperators in the network make up at least 80\%
of the population, over the long run.  This is considered a validation
of $b/c > \langle k \rangle$, as a necessary condition
for the establishment of cooperation in a social network \cite{ohtsuki}.  Such
validation against empirical data \cite{traud} is the main contribution of this study.

\section{Discussion} \label{dscssn}

Some of the very first formulations of the theory of games surfaced
during the first half of the twentieth century, when von Neumann
and Morgenstern \cite{neumort}, followed by Nash \cite{nash}, seeded 
foundations for a new field of study.

Prisoner's dilemma was invented by Merrill Flood and Melvin Dresher 
at the RAND corporation in 1950 \cite{nowaksup}.  Although 
its original formulation came from the point of view of classical
game theory, that is, with well-mixed populations.  Consideration 
of population structure in Prisoner's Dilemma was first conveyed with lattices or regular
networks.  For example, Nowak and May \cite{nowakmay92} proposed
a purely deterministic version of Prisoner's Dilemma on a two dimensional lattice.  This led
to a system that was extremely sensitive to initial conditions giving rise to fluctuations
in the densities of cooperators and defectors on the lattice.  In other words,
their system supports spatial arrays that vary chaotically, having
cooperation and defection shift in their sustained patterns \cite{nowakmay92}.

Regular lattices are often a good first approach while extending
a dynamical model to incorporate space.  However, when the structure in the population
is determined by social interactions, such as those maintained by players of an evolutionary game,
these regular graphs are limited descriptions.  The role of social structure is better
addressed by employing small-world networks \cite{abr1,ahmed00,cassar,chen08,chenwang08,deng10,dong10,du09,egui,
fu07,guan06, hauert05,kim02,masuda,perc06,szavuk,tang06,tomo04,thibert07,wu05,wu06,xia10,zhong06},
heterogeneous networks \cite{lenaerts,ohtsuki}, and datasets of real networks \cite{fu07,traud,traud12}.

There is a continued interest in exploring Prisoner's Dilemma on 
social networks with small-world properties.  In their pioneer 
introduction to small-world networks, Watts \& Strogatz \cite{wattstro}   
argued that as the fraction of rewired edges is increased,
then it is less likely for cooperation to emerge (with a Tit-for-Tat updating rule).  Moreover,
Watts \cite{watts99} explains that networks with very shy levels of clustering
tend to not enhance cooperation.  Because the establishment of cooperation
requires a critical mass of cooperators orchestrating against defectors, so that
they optimize their fitness or payoff by cooperating with each other.  According to
Watts \cite{watts99}, network shortcuts
can enable a few defectors to breakthrough the seed of cooperators, leading to
the eventual halt of the once sustained cluster of cooperation. 

On the other hand, small-world networks tend to favor cooperation
under a regime known as strategy dynamics.  Strategy dynamics is an approach
in which an initial set of updating rules are assigned in the first round, and for the following
rounds players may choose to switch between, say for example, Generalized Tit-for-Tat
and Copycat \cite{watts99}.  

For over a decade, efforts in exploring Prisoner's Dilemma on small-world networks footprints
a growing literature.  Here we comment on what we consider key citations, but 
we invite the reader to consult an extended list of references 
\cite{abr1,ahmed00,cassar,chen08,chenwang08,deng10,dong10,du09,egui,
fu07,guan06, hauert05,kim02,masuda,perc06,szavuk,tang06,tomo04,thibert07,wu05,wu06,xia10,zhong06} 
and references therein.

Even though several variations of Prisoner's Dilemma (a common approach is to
re-parametrize the payoff matrix, resulting in a matrix with only one parameter called
the temptation to defect) and its
 updating rule are considered, a distinct
consistent message is prevalent:  cooperation can persist in small-world networks.

For example, Abramson and Kuperman \cite{abr1} argue that in small-world networks
with an average degree of four, compact groups of
cooperators are seen to persist.  Moreover, long range edges, by means of moderate values
of the rewiring probability, favor cooperators as they start to reconnect, thus outcompeting 
defectors \cite{abr1}.  Tomochi \cite{tomo04}  discusses how random connections (rewiring)
enable breakthroughs of cooperation among clusters of defectors, leading to an unexpected 
scenario, where niches of defectors form and do not have incentives to switch their strategy, 
thus imposing over cooperators.  Hauert and Szab\'{o} \cite{hauert05} use the ratio of 
cost to net benefit of cooperation as a parameter while
exploring phase transitions, between cooperation and defection, in models
with network structure.  Furthermore, clusters of cooperators persist
with diffusion, that relocates these cooperators to other sites in a square lattice.  
Hauert and Szab\'{o} \cite{hauert05} also note regular small-world networks are even
more favorable to sustained cooperation than square lattices.  Perc \cite{perc06}
addresses the effects of extrinsic stochastic payoff functions, considered
as spatio-temporal random variations in Prisoner's Dilemma.  Additionally, Perc \cite{perc06}
finds an optimal fraction of rewired edges supports noise-induced cooperation with
resonance.  Xia, et al. \cite{xia10} employ co-evolutionary small-world networks
in a Prisoner's Dilemma game and they find that social structure collapses with
avalanches, by attacking the best cooperator hubs.  They argue that mutation
of the wealthiest (as determined by payoff) cooperators may promote sustained
cooperation on a large scale \cite{xia10}.

Prisoner's dilemma and social networks have been studied using
samples of real data.  Fu, et al. \cite{fu07}, analyze a dataset sampled from a Chinese
social networking site, which it is dubbed the Xiaonei dataset.  First, they compute
the clustering coefficient and characteristic path length, and conclude this dataset
has small-world properties.  Second, Fu, et al. \cite{fu07},  explain that
the evolution of cooperation in a Xiaonei dataset, is influenced by several social network
attributes, including: average connectivity, small-world effect, and degree-degree correlations.
Their numerical simulations of Prisoner's Dilemma on the Xiaonei dataset suggest
cooperation is substantially promoted, whenever the temptation-to-defect parameter
remains bounded, between 1.00 and 1.35.

The contribution by Fu, et al. \cite{fu07}, shares similarities with this study.  Because
here we also employ a dataset sampled from a social networking site along with
simulations of Prisoner's Dilemma.  


This study was inspired mainly by the contributions of Ohtsuki, et al. \cite{ohtsuki}, and Fu, et al. \cite{fu07}.  The former
conveys the cooperation probability of fixation.  That is, the probability that a single cooperator, located in a random
node of the network, in fact, converts the entire population from defectors into cooperators.  A network of size $n$, according
to \cite{ohtsuki}, has defectors with a fixation probability below $1/n$ and it has cooperators with
a fixation probability above $1/n$, provided that {\em ratio of benefit to cost
exceeds the average connectivity}.  In symbols,
we write $b/c > \langle k \rangle$ and note
this condition is necessary for cooperators to be favored by selection (this inequality is derived
by applying pair and diffusion approximations under the assumption that $n$ is considerably larger than $\langle k \rangle$).
Another interpretation of the discovery found by Ohtsuki, et al. \cite{ohtsuki}, is that natural selection promotes cooperation, 
with higher likelihood, when there are fewer connections.

On the other hand, Fu, et al. \cite{fu07}, analyzed a dataset of a real social network.  They
employed a sample of a friendship network, from a social networking site in China.  According
to their simulations of Prisoner's Dilemma, cooperation can reach as much as 80\% of the
network, for a range of values of the temptation to defect parameter.  Moreover,  Fu, et al. \cite{fu07},
argued that degree heterogeneity is fundamental for the establishment of cooperation
in friendship networks.

Here we have confirmed that cooperation may persist among social networks, provided some conditions are guaranteed.  
First, to draw a comparison, we simulate Prisoner's Dilemma on well-mixed populations and confirm that cooperation goes extinct regardless of
any values of benefit $b$ and cost $c$.  Then, to contrast the well-mixed scenario, 
we examine the persistence of cooperation with simulated social networks and with a dataset of a real social network.  
Prisoner's dilemma was studied in simulated networks between the two
extremes of small-world structures, that is, between regular graphs and random graphs, i.e., with rewiring
$p=0$ and $p=1$, respectively.  Cooperation keeps
sustained levels in both types of simulated social structures, with median levels of 80\% in regular graphs 
and 66\% in random graphs.  The skewness evidenced in the boxplots of the samples of cooperator density,
suggests that despite the fourteen percent drop in the median levels of sustained cooperation, extinction is
not a common occurrence.  We must note that the simulations on well-mixed and small-world populations were carried out
using the same game parameter values: $b=1.8$, $c=0.3$ and $w=0.5$.  The average degree in the simulated networks
was set to $\langle k \rangle = 4$, which means that $b/c = 6 > 4 = \langle k \rangle$.

Furthermore, cooperation persists among a real social network.  The latter determined
by a snapshot sample of a friendship network, in a collegiate social networking site, 
during its early days when there were domain restrictions for members \cite{traud}. Simulations
evidencing cooperation persistence were carried out with parameter values that 
satisfied the condition $b/c > \langle k \rangle$.  This serves as a
validation of the main result by Ohtsuki, et al. \cite{ohtsuki}, against a dataset of a real social network.
In fact, the median of sustained cooperation reaches 84\% of the social network.  Albeit some variability,
it is clear that cooperation among the facebook friendship network explored here draws a substantial
contrast with a well-mixed population.

We end with a note on further potential future directions of social network analysis and game theory.  More and more the 
field of mathematical epidemiology is integrating techniques
from evolutionary game theory, in the context of vaccination and behavioral changes  \cite{bauch03,chen12,shim11}.  For example, 
those vaccinating on-time can be considered cooperators, while those who do not vaccinate
can obtain the benefit of heard immunity, and may be considered defectors.  Studies involving
datasets of real social networks can shed some new light, when considering
a game theoretic approach to control epidemics.


\section*{Acknowledgements}
S. Cameron was funded by Talent Expansion in Quantitative Biology program 
(National Science Foundation grant  DUE-0525447) to attend
a two-day undergraduate workshop held at the Statistical and Applied Mathematical Sciences Institute (SAMSI), October 29--30, 2010.
S.M Cameron also received funding through a Research Discovery position given by ETSU Honors College, Summer 2011.
Contributions to this work were made while A. Cintron-Arias was visiting SAMSI, these visits were sponsored 
by East Tennessee State University Presidential-Grant-in-Aid E25150, and by SAMSI Working Group Dynamics On Networks.

\section*{Appendix: Simulation of Prisoner's Dilemma on a Social Network}
The initial conditions are the following.  Suppose a network with $n$ nodes is used to simulate the Prisoner's Dilemma.
An integer number $m$, such that $1\leq m \leq n$, is sampled uniformly at random from $(1,n)$.  Thus, $m$ nodes are selected uniformly
at random in the network and are set with strategy $D$, while all the other ones are set with strategy $C$.

\begin{enumerate}
	\item Choose one dying node uniformly at random, say it is node $i$.
	\item Compute the neighborhood of the dying node, say $\Omega_i$.
	\item Compute the payoff and fitness of every node $j\in\Omega_i$.
	\item Compute the aggregate fitness in $\Omega_i$ for each strategy:
		\begin{enumerate}
			\item aggregate fitness of all $C$-players in $\Omega_i$, say $F_C$.
			\item aggregate fitness of all $D$-players in $\Omega_i$, say $F_D$.
		\end{enumerate}
	\item	Let the empty site (dying node) adopt a strategy proportional to aggregate fitness.  Suppose
	$\alpha = \min(F_C,F_D)$ and $\beta = \max(F_C,F_D)$.  Consider
	the following cases.
	\begin{enumerate}
		\item Case 1: $\alpha,\beta>0$.  Sample $y\sim\mbox{Uniform}(0,\alpha +\beta)$.  If $y\leq\alpha$ then the empty site
		adopts the strategy associated with $\alpha$, i.e., it adopts $C$ if $\alpha=F_C$ or $D$ if $\alpha=F_D$.  Otherwise
		the dying node adopts the strategy associated with $\beta$.
		\item Case 2: $\alpha,\beta<0$.  Sample $y\sim\mbox{Uniform}(\alpha+\beta,0)$.  If $\beta\leq y <0$ then
		the empty site adopts the strategy associated with $\beta$.  Otherwise it adopts the strategy associated with $\alpha$.
		\item Case 3: $\alpha<0$ and $\beta>0$.  Sample $y\sim \mbox{Uniform}(\alpha,\beta)$.  If $\alpha\leq y<0$, then
		the dying node adopts the strategy associated with $\alpha$.  Otherwise it adopts the strategy associated 
		with $\beta$.
	\end{enumerate}

\end{enumerate}

\end{document}